\documentclass[twocolumn,preprintnumbers,amsmath,amssymb]{revtex4}
\usepackage{graphicx}
\usepackage{dcolumn}
\usepackage{bm}
\newcommand{\comment}[1]{}
\newcommand\etal{\mbox{\textit{et al.}}}

\begin{document}

\preprint{}

\title{Dynamics of spectrally truncated inviscid turbulence}
\author{W.J.T. Bos (correspondent)}
\author{J.-P. Bertoglio}

\affiliation{%
 LMFA, UMR CNRS 5509\\ Ecole Centrale de Lyon - Universit\'e Claude Bernard Lyon I - INSA Lyon\\ 69134 Ecully Cedex, France}

\begin{abstract}The evolution of the turbulent energy spectrum for the inviscid spectrally truncated Euler equations is studied by closure calculations. The observed behavior is similar to the one found in Direct Numerical Simulations (Cichowlas, Bona\"{\i}titi, Debbasch and Brachet [Phys. Rev. Lett.  {\bf 95}: 264502, 2005]). A Kolmogorov spectral range and an equipartition range are observed simultaneously. Between these two ranges a ``quasi-dissipative'' zone is present in the kinetic energy spectrum. The time evolution of the wavenumber that marks the beginning of the equipartition range is analyzed and it is shown that spectral nonlocal interactions are governing this evolution.
\end{abstract}

\maketitle

Recently renewed interest was shown in the evolution of inviscid turbulence. Cichowlas, Bona\"{\i}titi, Debbasch and Brachet \cite{Cichowlas} (in the following abbreviated as CBDB), reported results of Direct Numerical Simulations (DNS) of the spectrally truncated 3-D incompressible Euler equations, with resolutions of $256^3$, $512^3$, $1024^3$ and $1600^3$ wave-modes. In principle, the presence of a high-frequency spectral truncation allows the flow to reach a thermal equilibrium, i.e. an equipartition of energy over wave-vectors. The spherically averaged energy spectrum $E(K)$ associated to this equilibrium has a $K^2$ wavenumber dependence. This is effectively observed in the CBDB DNS, where at high wavenumbers the energy spectrum is found to increase following a $K^2$ dependence. It was also observed in CBDB that during the transient towards the equilibrium, $E(K)$ followed a $K^{-5/3}$ Kolmogorov scaling in the intermediate range. In between these two power law regions, a ``quasi-dissipative'' zone is found in which the energy spectrum falls off faster than $K^{-5/3}$ until a local minimum of the spectrum is reached. 

The existence of a $K^2$ range in the spectrum is not new and equipartition was discussed long ago in the literature. \cite{Kraichnan67,Montgomery} The observation of a Kolmogorov scaling in the energy spectrum of the truncated Euler equations is more interesting as it shows that viscous dissipation is not mandatory for the build up of an inertial zone but that a flux of energy to the small scales, where it accumulates, is sufficient. It is also an interesting observation that between the two coexisting power laws, a dip or quasi-dissipative zone, is detected.

The Eddy-Damped Quasi-Normal Markovian theory (EDQNM) \cite{Orszag,Leith2} is known to be compatible with both the equipartition of kinetic energy \cite{Montgomery,Carnevale} and the existence of a $K^{-5/3}$ inertial range. Calculations at higher resolution than DNS can be performed at a much lower computational cost. EDQNM therefore appears as an adequate tool to investigate the high-resolution spectral dynamics of the incompressible Euler equations. 

In the present study it is first shown that the spectral behavior observed in DNS can be reproduced satisfactorily by this relatively simple statistical closure. Results for resolutions currently unattainable by DNS are analyzed. Eventually, it is shown that the dynamics of the ``quasi-dissipative'' zone are governed by highly nonlocal triad interactions.

In the case of the EDQNM closure, the evolution equation for the energy spectrum corresponding to the Euler equations is the Lin equation without viscosity:
\begin{equation}
\frac{\partial E(K,t)}{\partial t}=T_{NL}(K,t)
\end{equation}
in which the nonlinear transfer $T_{NL}$ is expressed as:
\begin{eqnarray}\label{TNL}
T_{NL}(K,t)=\iint_{\Delta}\Theta_{KPQ}~(xy+z^3)\left[K^2PE(P,t)E(Q,t) \right.\nonumber\\
\left. -P^3E(Q,t)E(K,t)\right]\frac{dPdQ}{PQ}~~~~~~~~~
\end{eqnarray}
In equation (\ref{TNL}), $\Delta$ is a band in $P,Q$-space so that the three wave-vectors ${\bm{K}, \bm{P}, \bm{Q}}$ form a triangle. $x,y,z$ are the cosines of the angles opposite to $K,P,Q$ in this triangle. The characteristic time $\Theta_{KPQ}$ is defined as:

\begin{equation}
\Theta_{KPQ}=\frac{1-exp(-(\eta_K+\eta_P+\eta_Q)\times t)}{\eta_K+\eta_P+\eta_Q}
\end{equation}
in which $\eta$ is the eddy damping, expressed as
\begin{equation}
\eta_K=\lambda\sqrt{\int_0^K S^2E(S,t)dS}.
\end{equation}
For $\lambda$ the classical value $0.36$ is retained corresponding to a value for the Kolmogorov constant of $1.4$. The spatial resolution for the computations reported here is $14.2$ wavenumbers per octave. This significantly  high wavenumber density was selected in order to satisfactorily capture the nonlocal interactions that will be shown to play an important role in the following. The effect of the truncation of the domain at a cut-off wavenumber $K_{max}$ is  introduced by omitting in (\ref{TNL}) all interactions involving wavenumbers larger than $K_{max}$.

In the present paper we analyze a freely evolving inviscid velocity field (the term decaying would be misleading since no energy is dissipated). The initial spectral energy distribution is localized at small wavenumbers. The initial energy spectrum is: 
\begin{equation}\label{initexp}
E(K,0)=B K^se^{-2K^2/K_L^2},
\end{equation}
with $s=4$ and $B$ determined so that the total kinetic energy is equal to unity. In this spectrum $K_L$ is a wavenumber characterizing the initial integral length scale, and the computations are run on a spectral domain extending from a minimum wavenumber $K_0$ to the high frequency cut-off $K_{max}$.  These initial conditions are different from the ones used by CDBD, who started from a Taylor Green vortex, so that the short time evolutions may be different in the two studies. A first set of computations was performed with $K_0=K_L$. In this case the maximum of the energy spectrum is situated at the first wavenumber and the $K^s$ part of the spectrum is absent. This set has the advantage of permitting comparisons with the DNS of CBDB in which the energy maximum is also located at the low wavenumbers at the beginning of the computations. Calculations are performed with ratios of smallest to largest wavenumbers $K_{max}/K_0=85,~171,~341,~533$. These resolutions are equivalent to the $256^3$, $512^3$, $1024^3$ and $1600^3$ DNS calculations of CBDB respectively. Calculations were also performed at higher resolutions, that would correspond to DNS on $4096^3$, $8192^3$, $16384^3$ and $32768^3$ grids. In the following the different calculations will be denoted by $256$, $512$, etc. Finally we report the results of a calculation starting with a spectrum having a $K^4$ low wavenumber zone. In this case $K_L/K_0=10$ and $K_{max}/K_0=10^4$. In the following the time is normalized by an initial eddy turnover time defined by $(\mathcal U K_L)^{-1}$, in which $\mathcal U=\left(\frac{2}{3}\int E(K)dK\right)^{1/2}$.

\begin{figure} 
\begin{center}
\setlength{\unitlength}{1.\textwidth}
\includegraphics[width=0.5\unitlength]{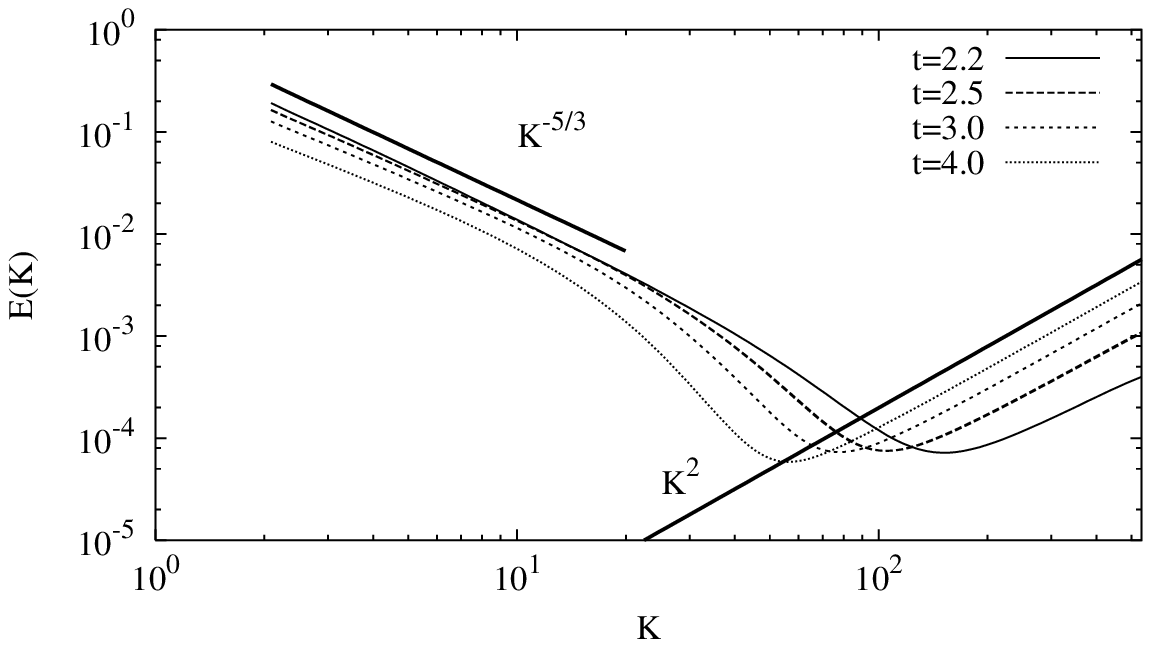}
\includegraphics[width=0.5\unitlength]{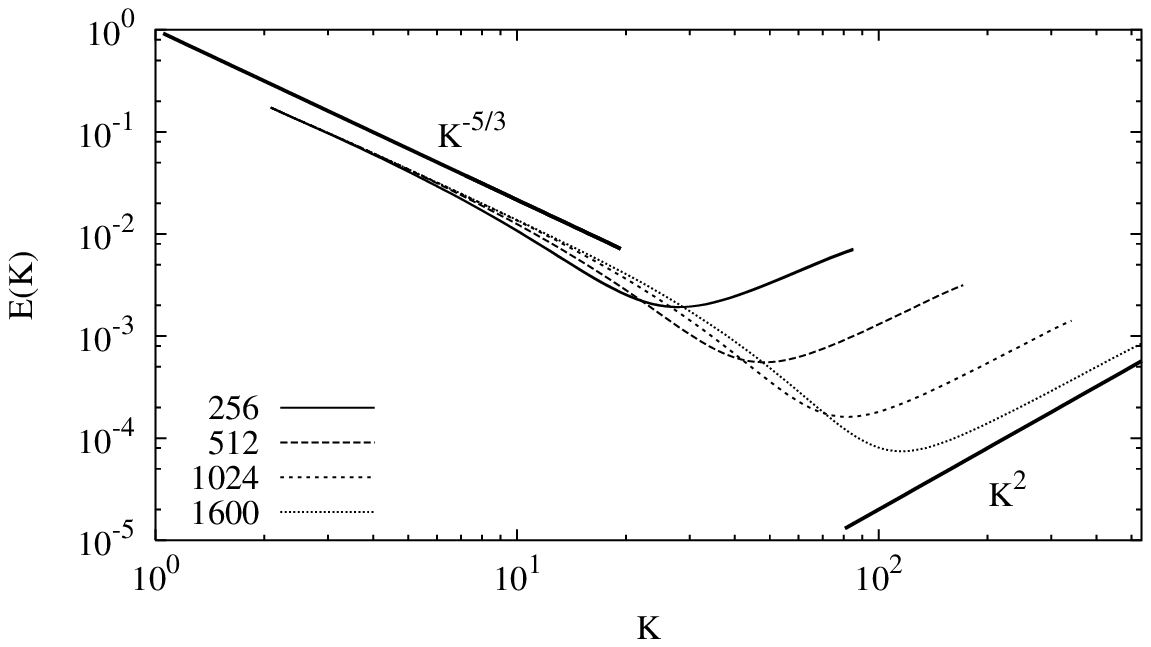}
\includegraphics[width=0.5\unitlength]{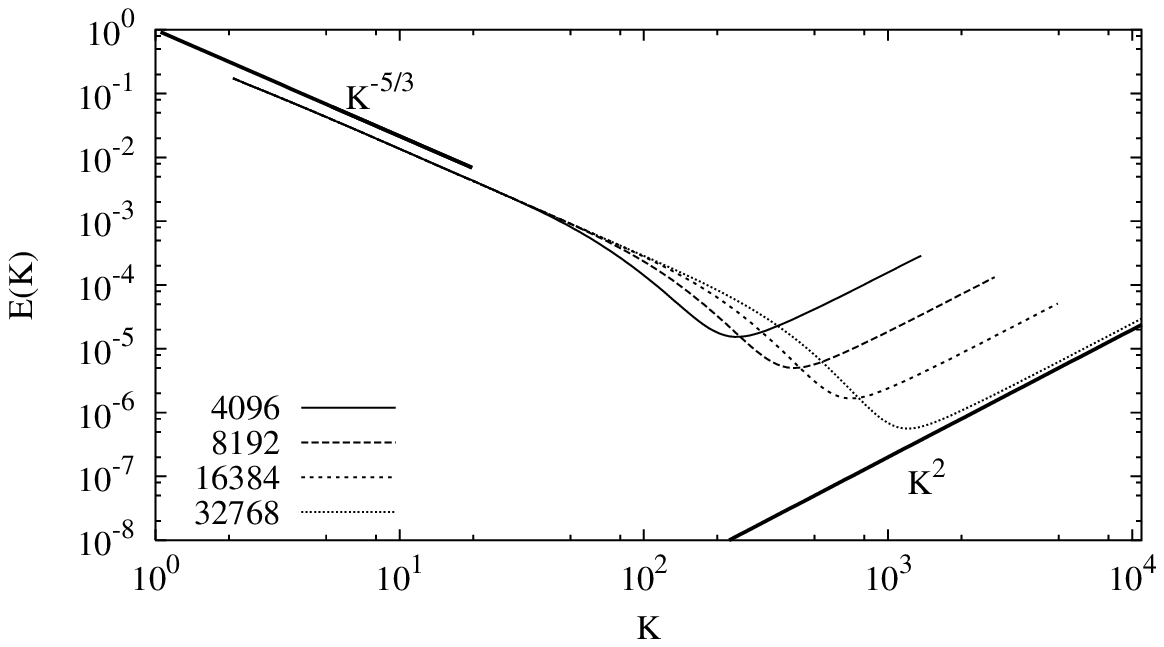}
\end{center}
\caption{Top: time evolution of the energy spectrum $E(K)$ for calculation $1600$. The solid lines indicate the $K^{-5/3}$ and $K^2$ spectral slopes. Middle: spectra for the calculations $256$, $512$, $1024$ and $1600$ at $t=2.4$. Bottom: spectra for the calculations $4096$, $8192$, $16384$ and $32768$ at $t=2.4$.
}
\end{figure}

\begin{figure} 
\begin{center}
\setlength{\unitlength}{1.\textwidth}
\includegraphics[width=0.5\unitlength]{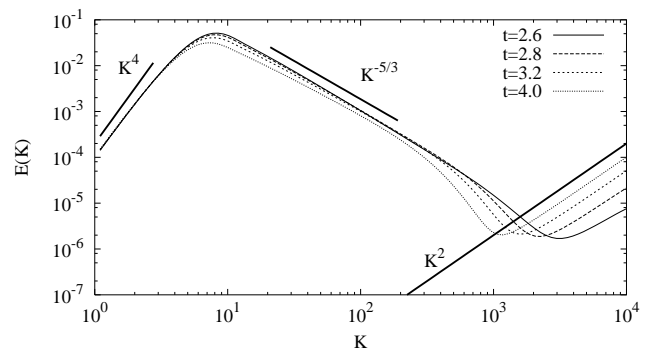}
\end{center}
\caption{Time evolution of an energy spectrum with a $K^4$ low wavenumber range.}
\end{figure}

The results for the time evolution of the $1600$ calculation are shown in figure 1. Also shown are the spectra at $t=2.4$ for different resolutions. Comparison with Figure 1 of reference \cite{Cichowlas} shows that the present closure results are very similar to the DNS results. The results of the calculation starting from an initial spectrum with a $K^4$ low wavenumber range are shown in figure 2.

In all calculations the nonlinear interactions create a $K^{-5/3}$ inertial range. The existence of the inertial range is particularly clear in the results at high resolutions. 
In this range the energy spectrum scales as:
\begin{equation}\label{Kolmospec}
E(K,t)=C_K \epsilon(t)^{2/3}K^{-5/3},
\end{equation}
in which $\epsilon(t)$ should not be interpreted as a molecular dissipation which would be zero in the present case, but as an energy flux in the spectral cascade. This range elongates with time towards higher wavenumbers until the end of the inertial range meets the maximum wavenumber of the truncated domain $K_{max}$. Then a thermalized range, or equipartition spectrum,
\begin{equation}\label{AK2}
E(K,t)=A(t) K^2, 
\end{equation}
starts to build up at large wavenumbers. This thermalized range later extends from large to smaller wavenumbers, and $A(t)$ is an increasing function of time. The thermalized energy $e_{th}$ associated with this range is also increasing with time. By integrating the equipartition range it is found that $A(t)$ is related to $e_{th}$ by the expression $A(t)\approx 3e_{th}/K_{max}^3$. For long times the thermalized energy approaches the (invariant) total energy and $A(t)$ tends to a constant value.

The coexistence of a Kolmogorov range and of an equipartition zone was already observed by Connaughton and Nazarenko \cite{Connaughton} using the diffusion approximation of turbulence proposed by Leith \cite{Leith}. The ``quasi-dissipative'' zone is however absent in their results. In agreement with the DNS results of CBDB, the "dissipative" range is present in the EDQNM results. Leith's model was shown to be a useful and easily maneagable model to obtain understanding of various problems related to turbulence (e.g. scalar mixing, two-dimensional turbulence \cite{Leith3,Leith4}). Its character is however spectrally local. EDQNM, on the contrary, basically relies on the triadic interaction between wave-modes and is therefore by essence nonlocal. The fact that the "dissipative" range is reproduced by the EDQNM closure, while it is absent in the results obtained with Leith's diffusion model \cite{Connaughton}, indicates that the nonlocal interactions between the thermalized modes and the modes in desequilibrium might be responsible for the existence of this zone. This issue will now be addressed.

As in CBDB we introduce $K_{th}(t)$ as the wavenumber corresponding to the minimum of the energy spectrum. $K_{th}(t)$ characterizes the beginning of the absolute equilibrium zone. 
From the equations (\ref{Kolmospec}) and (\ref{AK2}), it is straightforward to calculate the intersection of the two zones, yielding a first estimate for $K_{th}(t)$:
\begin{equation}\label{KminBR}
K_{pl}(t)\sim\left(\frac{\epsilon(t)}{e_{th}(t)^{3/2}}\right)^{2/11}K_{max}^{9/11},
\end{equation}
The subscript $pl$ indicates that a $K^{-5/3}$ power law inertial range is presumed reaching the equipartition range at $K=K_{th}(t)$. This estimate was first proposed in Cichowlas \etal \cite{CichowlasIUTAM2}. Another possible estimate is:
\begin{equation}\label{KminBR2}
K_{d}(t)\sim\left( \frac{\epsilon(t)}{ e_{th}(t)^{3/2}}\right)^{1/4}K_{max}^{3/4}.
\end{equation}
It was also proposed in reference \cite{CichowlasIUTAM2} (see also CBDB). The physical argument behind this estimation is the existence of a ``quasi-dissipative'' region in the spectrum, occurring between the Kolmogorov and equipartition ranges. A simple way to find $K_d$ by dimensional arguments is to postulate that $K_d$ is the inverse of a Kolmogorov scale
\begin{equation}
\frac{1}{K_d}\sim \left(\frac{\nu_t^3}{\epsilon}\right)^{1/4}
\end{equation}
built on the energy flux $\epsilon$ and an eddy viscosity $\nu_t$. Assuming $\nu_t$ determined by the thermalized energy and the wavenumber bound of the spectral domain $K_{max}$,
\begin{equation}\label{eddyvis}
\nu_t\sim \frac{\sqrt{e_{th}}}{K_{max}},
\end{equation}
(\ref{KminBR2}) is immediately found. Introducing $K_{max}^{-1}$ as the lengthscale on which to build the effective viscosity $\nu_t$ is an assumption typically nonlocal in wave-space. 

The nonlocal nature of estimate (\ref{KminBR2}) can be enlightened using the EDQNM equations and performing a nonlocal expansion {\it \`a la} Lesieur and Schertzer \cite{Lesieur78}. Developing the nonlinear transfer (\ref{TNL}) with respect to $K/K_i$ and assuming $K<< K_i$ and $E(K)< E(K_i)$ yields:

\begin{eqnarray}
T_{NL}(K)=
-2\nu_tK^2E(K)\nonumber\\
+\frac{4}{15}K^4\int_{K_i}^{K_{max}}\Theta_{KPP}\frac{E(P)^2}{P^2}dP 
+\mathcal{O}\left[KE(K)\right]^{3/2}~~~
\end{eqnarray}
Focusing on the first term that has the form of an eddy viscous term with:
\begin{eqnarray}
\nu_t=\frac{1}{15}\int_{K_i}^{K_{max}}\Theta_{KPP}\left[5 E(P) +P\frac{\partial E(P)}{\partial P}\right]dP\nonumber\\
\end{eqnarray}
and assuming $E(K_i)$ in the equipartition range, it is readily found that $\nu_t$ is given by expression (\ref{eddyvis}). The fact that expression (\ref{eddyvis}), and therefore estimate (9), can be found by a non local expansion of the tranfer term, underlines the non local nature of the physical mechanisms associated with this estimate. It has to be noted that expression (\ref{KminBR2}) was proposed by CBDB following a different argument.

\begin{figure} 
\begin{center}
\setlength{\unitlength}{1.\textwidth}
\includegraphics[width=0.5\unitlength]{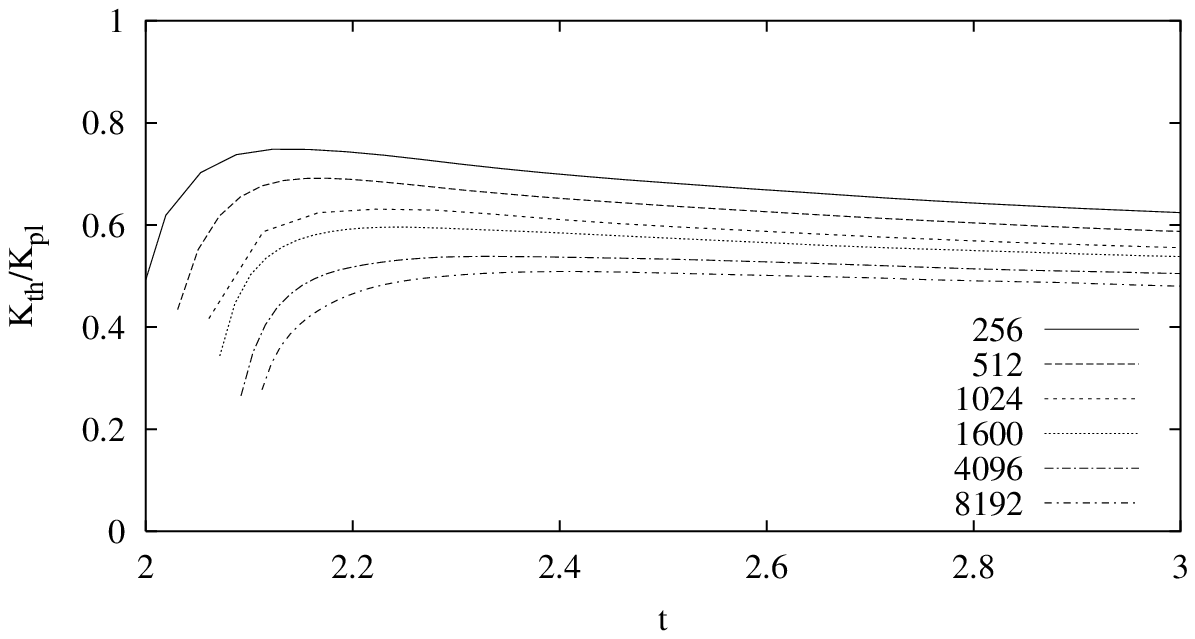}
\includegraphics[width=0.5\unitlength]{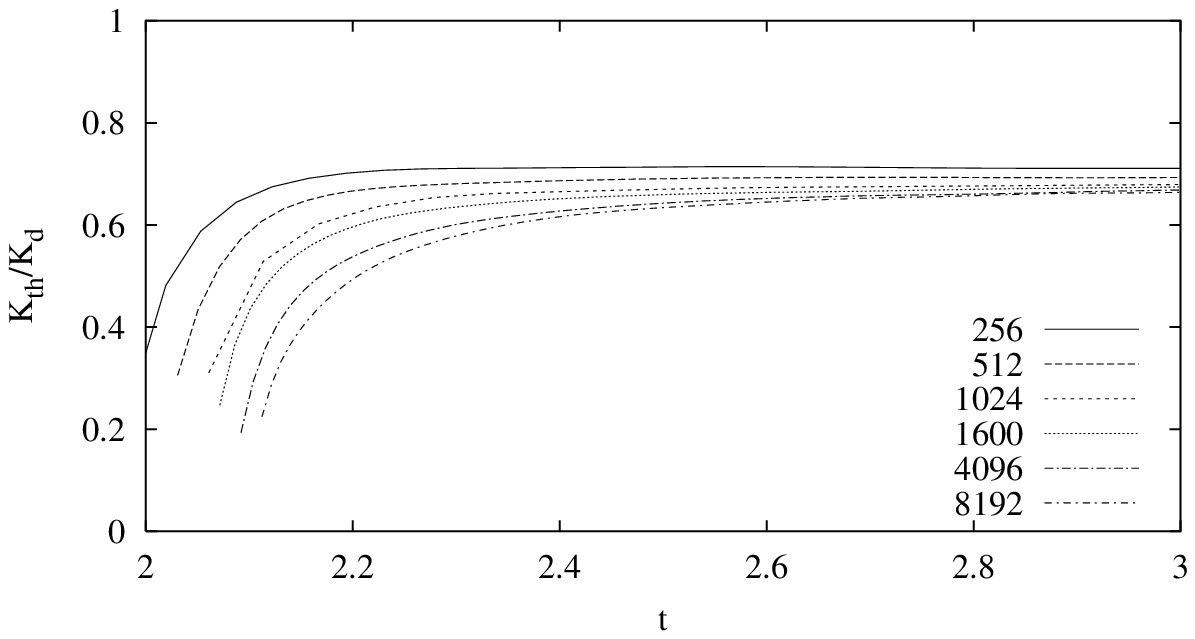}
\end{center}
\caption{Evolution of the parameter $K_{th}(t)$. Top: $K_{th}$ compared to $K_{pl}$ (equation (\ref{KminBR})). Bottom: $K_{th}$ compared to $K_{d}$ (equation (\ref{KminBR2})).}
\end{figure}

\begin{figure} 
\begin{center}
\setlength{\unitlength}{1.\textwidth}
\includegraphics[width=0.5\unitlength]{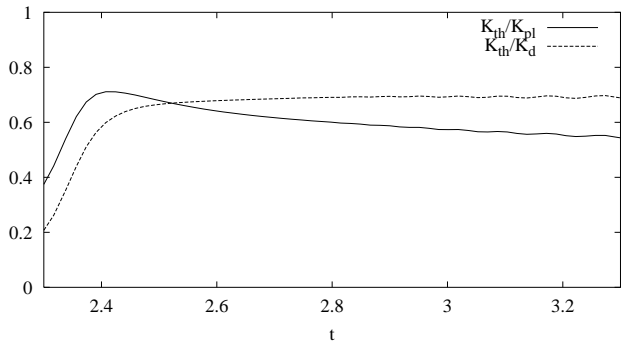}
\end{center}
\caption{Evolution of $K_{th}(t)$, compared to the two estimates $K_{pl}(t)$ and $K_{d}(t)$ (equations (\ref{KminBR}) and (\ref{KminBR2}) respectively) for the $K^4$ case.
}
\end{figure}

While in DNS it is not easy to decide which estimator for $K_{th}$ is the more appropriate, the relative cheapness of EDQNM calculations allows for high resolution calculations that help to choose between the two estimates. In figure 3 the values of $K_{th}$ found with the closure are compared to the estimates (\ref{KminBR}) and (\ref{KminBR2}). The dissipative estimate is clearly found to be the more appropiate since it leads to a better collapse of the results on the different runs as well as to a more pronounced horizontal plateau of the plotted ratios. In figure 4 it is confirmed that also in the $K^4$ case the dissipative estimate of $K_{th}(t)$ is superior to the power-law estimate. It is interesting to point out here that the opposite conclusion would be obtained with a model  that excludes the nonlocal interactions as in Connaughton and Nazarenko \cite{Connaughton}. 

The main results of the study can be summarized as follows. The EDQNM closure reproduces the behavior observed in DNS, i.e. an energy spectrum containing a Kolmogorov inertial range, a dissipation range and an equipartition range. The dissipation range was shown to be created by nonlinear interactions with the modes in equipartition. An effective eddy viscosity can be defined, based on the most energetic modes in the equipartition zone and the cut-off wavenumber. The non local character of this eddy viscosity was verified by expanding the non-local contributions to the nonlinear transfer in the EDQNM formulation as a function of the wavenumber ratio and retaining the leading order term. 

Although the problem of truncated inviscid turbulence remains somewhat artificial and far from real world applications, some of the spectral behaviors studied here may help understanding mechanisms that are connected to Navier-Stokes turbulence. It is for example interesting to note, in the framework of Large-Eddy Simulation, that the turbulent cascade of energy at large scales is not significantly altered by the fact that energy is not, or not properly, dissipated. 

We acknowledge stimulating interaction with Marc-Etienne Brachet in the framework of the GDR  ``Structure de la Turbulence et M\'elange'' of the Centre National de la Recherche Scientifique (CNRS), which provided the starting point for the present work. 

\end{document}